\documentclass[12pt]{article}

\usepackage{graphicx}

\def\beq   {\begin{equation}}
\def\eeq   {\end{equation}}
\def\beqd  {\begin{displaymath}}
\def\eeqd  {\end{displaymath}}
\def\bea {\begin{eqnarray}}
\def\eea {\end{eqnarray}}

\def\psla{p\kern-.45em/}
\def\qsla{q\kern-.45em/}
\def\ksla{k\kern-.45em/}
\def\dels{\partial\kern-.45em/}

\def\st{\ifmmode{\tilde{t}} \else{$\tilde{t}$} \fi}
\def\sb{\ifmmode{\tilde{b}} \else{$\tilde{b}$} \fi}
\def\sq{\ifmmode{\tilde{q}} \else{$\tilde{q}$} \fi}
\def\sg{\ifmmode{\tilde{g}} \else{$\tilde{g}$} \fi}

\def\sz{\ifmmode{\tilde{\chi}^0} \else{$\tilde{\chi}^0$} \fi}
\def\sw{\ifmmode{\tilde{\chi}} \else{$\tilde{\chi}$} \fi}

\def\dr{\ifmmode{\overline{\rm DR}} \else{$\overline{\rm DR}$} \fi}
\def\ms{\ifmmode{\overline{\rm MS}} \else{$\overline{\rm MS}$} \fi}

\begin{document}

\pagestyle{empty}

\vspace*{-1cm} 
\begin{flushright}
  TU-612 \\
  hep-ph/0103046
\end{flushright}

\vspace*{1.4cm}

\begin{center}

{\Large {\bf
Gauge dependence of the on-shell renormalized mixing matrices
} 
}\\

\vspace{10mm}

{\large Youichi Yamada}

\vspace{6mm}
\begin{tabular}{l}
{\it Department of Physics, Tohoku University, Sendai 980-8578, Japan}
\end{tabular}

\end{center}

\vspace{3cm}

\begin{abstract}
It was recently pointed out that the on-shell renormalization of 
the Cabibbo-Kobayashi-Maskawa (CKM) matrix in the method 
by Denner and Sack causes a gauge parameter dependence of the amplitudes. 
We analyze the gauge dependence of the on-shell renormalization of 
the mixing matrices both for fermions and scalars in general cases, 
at the one-loop level. 
We then show that this gauge dependence can be avoided by 
fixing the counterterms for the mixing matrices in terms of 
the off-diagonal wave function corrections for fermions and scalars 
after a rearrangement, in a similar manner to the 
pinch technique for gauge bosons. We finally present explicit calculation 
of the gauge dependence for two cases: the CKM matrix in the Standard Model, 
and left-right mixing of scalar quarks in the 
minimal supersymmetric standard model. 
\end{abstract}

\newpage
\pagestyle{plain}
\setcounter{page}{2}

\section{Introduction} 

Particles in the same representation under unbroken symmetries 
can mix with each other. The neutral gauge bosons, quarks, and 
massive neutrinos in the Standard Model (SM) are well-known examples. 
New particles in extensions of the Standard Model also 
show the mixings. For example, in the minimal supersymmetric (SUSY) 
standard model (MSSM) \cite{mssm}, a very promising extension, 
superpartners of most SM particles show the mixing \cite{mssm,gh}. 
The mixing of particles is expressed in terms of the mixing matrix, 
which represents the relations between the gauge eigenstates and the 
mass eigenstates of the particles. The mixing matrices 
always appear at the couplings of these particles in the mass eigenbasis. 

Because of the fact that mass eigenstates at the tree-level mix 
with each other by radiative corrections, 
the mixing matrices have to be renormalized \cite{sirlin,DS} to 
obtain ultraviolet (UV) finite amplitudes. 
Denner and Sack have proposed \cite{DS} a simple scheme to renormalize 
the mixing matrix of Dirac fermions at the one-loop level, which is 
usually called the on-shell renormalization scheme. 
They have required the counterterm for the renormalized mixing matrix 
to completely absorb the anti-hermitian part of the wave function 
correction $\delta Z_{ij}$ for the external on-shell fields. 
This definition works very well 
for the subtraction of the ultraviolet divergence and dependence on 
the renormalization scale. The renormalization procedure is universal 
for any processes with the particles as external states. It also absorbs the 
${\cal O}(1/(m_i^2-m_j^2))$ terms which are singular for the case 
$m_i\simeq m_j$. The on-shell scheme was also applied to the mixing of 
other fields, such as Majorana fermions \cite{KP} and complex scalar 
particles \cite{sqyukawa}. 

However, it has recently been pointed out \cite{gaugedep,KMS,barroso} that 
in the on-shell scheme of Ref.~\cite{DS} 
the counterterms for the Cabibbo-Kobayashi-Maskawa (CKM) 
matrix \cite{CKM} is dependent on the 
gauge fixing parameter and that, as a consequence, 
the amplitudes of charged current interactions of quarks are also 
gauge dependent in this scheme. 
This fact motivated these authors to introduce other ways for the 
UV finite renormalization of the CKM matrix \cite{gaugedep,barroso}. 
However, their method cannot be directly applied to 
mixings of other particles. 

In this paper we study the gauge parameter dependence of the 
on-shell renormalized mixing matrices in general cases. 
We demonstrate that this gauge dependence is a general 
feature for the on-shell mixing matrices. Nevertheless, 
at the one-loop level the on-shell mixing 
matrices by Ref.~\cite{DS} can be modified to be gauge independent by the 
following procedure. First, we split the gauge-dependent parts of the 
wave function corrections in the similar way 
to the ``pinch technique'' \cite{pinchold,pinch,pinchfermi}. 
They are then rearranged into the corresponding vertex corrections 
in the amplitudes and cancelled. 
Next, we give the counterterm for the 
on-shell renormalized mixing matrices in terms of the remaining, 
gauge-independent part of $\delta Z_{ij}$. The subtraction of the UV 
divergence and of the ${\cal O}(1/(m_i^2-m_j^2))$ singularity is 
not affected by this modification. This method can be applied 
in a similar manner both for mixings of fermions and of scalars. 

This paper is organized as follows. In section 2 we review the 
one-loop on-shell renormalization of the mixing matrices for 
scalars and fermions in general case. 
In section 3 their gauge dependences are analyzed by using the 
Nielsen identities \cite{nielsen1,PS,nielsenSM} for self energies of 
scalars and fermions. We then show that the 
gauge dependences of the off-diagonal wave function corrections and, 
in consequence, of the on-shell mixing matrices 
can be split by the rearrangement of the loop corrections. 
Sections 4 and 5 present two explicit calculations of the 
gauge dependence of mixing matrices; CKM matrix of quarks in the 
SM and left-right mixing of scalar quarks (squarks) in the MSSM. 
Section 6 gives our conclusion. 

\section{On-shell renormalization of mixing matrices}

Let $\psi_{\alpha}$ (with index $\alpha$) be fields in gauge eigenstates, 
either real or complex scalars, or chiral components of Dirac or 
Majorana fermions. 
The fields in common representation under unbroken symmetries 
may mix with each other to form mass eigenstates. 
The relation between gauge eigenstates $\psi_{\alpha}$ and tree-level 
mass eigenstates $f_i$ with masses $m_i$ is expressed 
by an unitary matrix $U$ as 
\beq
f_i= U_{i\alpha}\psi_{\alpha},\;\;\;
\psi_{\alpha}= U^*_{i\alpha}f_i.     \label{eq1}
\eeq
The mixing matrix $U$ is determined such that the tree-level mass 
matrix for $f_i$ is diagonal. The couplings of $f_i$ are always 
multiplied by $U$. For example, an amplitude ${\cal M}_i$ with 
one incoming external $f_i$ is expressed as 
\beq
{\cal M}_i = \sum_{\alpha}{\cal M}_{\alpha} U^*_{i\alpha}, \label{eq2}
\eeq
where ${\cal M}_{\alpha}$ has no $U$ dependence. $U$ is therefore 
very important parameter for $f_i$. Note that, when $f$ are fermions, 
the mixing matrices $U^L$ and $U^R$ for chiral 
components $f_L$ and $f_R$, respectively, are generally different from 
each other. 

By radiative corrections, the wave functions of $f_i$ should be 
renormalized. The on-shell renormalized fields $f_i$ are related 
to the unrenormalized $f_i^{(0)}$ by, at the one-loop level 
\beq
f_i^{(0)} = \left( \delta_{ij} + \frac{1}{2}\delta Z_{ij} \right) f_j . 
\label{eq3}
\eeq
The off-diagonal parts of $\delta Z_{ij}(i\neq j)$ represent 
the mixing between $f_i$ and $f_j$. 
For the relation (\ref{eq1}) is modified as 
\beq
\psi_{\alpha}= U^{(0)*}_{i\alpha}f^{(0)}_i = 
U^{(0)*}_{i\alpha}\left( \delta_{ij} + \frac{1}{2}\delta Z_{ij}\right)f_j, 
\eeq
the wave function correction to the amplitude (\ref{eq2}) is 
expressed as the replacement of $U$ by 
\beq
U^*_{i\alpha} \rightarrow 
U^{(0)*}_{j\alpha}\left( \delta_{ji} + \frac{1}{2}\delta Z_{ji}\right)\,. 
\label{eq5}
\eeq
This correction is universal in any processes involving 
on-shell external $f_i$. 

The explicit form of $\delta Z_{ij}$ for $i\neq j$ is given 
in terms of the off-diagonal, flavor-mixing parts of the 
self energy\footnote{We assume that 
the absorptive part of the self energy is negligible. 
For its correct inclusion one has to treat $f$'s as unstable 
intermediate states.} of the fields $f$.
For scalars with unrenormalized, dimensionally regularized 
self energy $\Pi_{ij}(p^2)$, we have 
\beq
\frac{1}{2}\delta Z_{ij} = \frac{1}{m_i^2-m_j^2}\Pi_{ij}(m_j^2). \label{eq6}
\eeq
For Dirac fermions with self energy 
\bea
&& \Sigma_{ij}(p)=\Sigma_{Lij}(p^2)\psla P_L+\Sigma_{Rij}(p^2)\psla P_R
+\Sigma_{DLij}(p^2)P_L+\Sigma_{DRij}(p^2)P_R~, \label{eq7} \\
&& \Sigma^*_{Lji}(p^2)=\Sigma_{Lij}(p^2),\;
\Sigma^*_{Rji}(p^2)=\Sigma_{Rij}(p^2),\;
\Sigma_{DRij}(p^2)=\Sigma^*_{DLji}(p^2), 
\eea
the corrections to chiral components of the 
wave functions ($f_{iL}$, $f_{iR}$) are \cite{onshell}
\begin{eqnarray}
\frac{1}{2}\delta Z^L_{ij}=\frac{1}{m_i^2-m_j^2}\left[
m_j^2\Sigma_{Lij}(m_j^2)+ m_im_j\Sigma_{Rij}(m_j^2) 
+m_i\Sigma_{DLij}(m_j^2)+m_j\Sigma_{DRij}(m_j^2) \right]~,
\nonumber\\
\frac{1}{2}\delta Z^R_{ij}=\frac{1}{m_i^2-m_j^2}\left[
m_im_j\Sigma_{Lij}(m_j^2)+ m_j^2\Sigma_{Rij}(m_j^2) 
+m_j\Sigma_{DLij}(m_j^2)+m_i\Sigma_{DRij}(m_j^2)
\right]~,  \label{eq9}
\end{eqnarray}
respectively. Both of Eqs.~(\ref{eq6},\ref{eq9}) have 
the factor $1/(m_i^2-m_j^2)$ which is unique for the off-diagonal 
wave function corrections. 
These $\delta Z_{ij}$ are UV divergent and depend on 
the gauge fixing parameters $\xi$ for the massive gauge bosons. 
Note also that $\delta Z_{ij}$ superficially diverge 
when the masses ($m_i$, $m_j$) of $f_i$ and $f_j$, respectively, 
become close to each other. 
For the case of Majorana fermions \cite{KP}, the self energy (\ref{eq7}) 
obeys additional conditions 
\begin{equation}
\Sigma_{Lij}(p^2)=\Sigma^*_{Rij}(p^2),\;
\Sigma_{DLij}(p^2)=\Sigma_{DLji}(p^2),\;
\Sigma_{DRij}(p^2)=\Sigma_{DRji}(p^2). \label{eqMajo}
\end{equation}
The condition for the wave function corrections, 
$\delta Z^L_{ij}=\delta Z^{R*}_{ij}$, which is necessary for keeping 
Majorana condition $U^L=U^{R*}$ after renormalization, then follows from 
Eqs.~(\ref{eq9}, \ref{eqMajo}). All subsequent discussions in this 
and the next sections remain unchanged by the conditions (\ref{eqMajo}). 

For the cancellation of the UV divergence of off-diagonal $\delta Z_{ij}$ 
in Eq.~(\ref{eq5}), the mixing matrix $U$ has to be 
renormalized \cite{sirlin,DS}. 
Assume that the renormalized $U$ is related to the bare $U^{(0)}$ by 
\beq
U^{(0)}_{i\alpha}=(\delta_{ij}+\delta u_{ij})U_{j\alpha}\, . \label{eq10}
\eeq
Since both $U^{(0)}$ and $U$ are unitary, the counterterm $\delta u$ should 
be anti-hermite. The correction factor (\ref{eq5}) is then rewritten as 
\beq
U^{(0)*}_{j\alpha}\left( \delta_{ji} + \frac{1}{2}\delta Z_{ji}\right) 
= U^*_{j\alpha}\left( \delta_{ji} + \frac{1}{2}\delta Z_{ji}
-\delta u_{ji} \right).  \label{eq11}
\eeq
The UV divergent part of $\delta u$ is determined \cite{DS} such as 
to cancel that of the anti-hermitian part of $\delta Z$. For fermions, 
also the UV divergence of the diagonal CP-violating part 
\beq
\frac{i}{2}{\rm Im}(\delta Z^L_{ii})= 
-\frac{i}{2}{\rm Im}(\delta Z^R_{ii})= 
\frac{i}{2m_i}{\rm Im}(\Sigma_{DLii}(m_i^2)),  \label{eq9a}
\eeq
in the convention\footnote{For Dirac fermions, one may make the shift 
($\delta Z^L_{ii}$, $\delta Z^R_{ii}$)$\rightarrow$
($\delta Z^L_{ii}+i\theta_i$, $\delta Z^R_{ii}+i\theta_i$) by 
an arbitrary imaginary number $i\theta_i$. 
This is equivalent to the phase rotation ($f_{iL}$, $f_{iR}$)$\rightarrow$
($e^{i\theta_i}f_{iL}$, $e^{i\theta_i}f_{iR}$) in Eq.~(\ref{eq3}). 
This freedom is killed by Majorana condition. 
See Ref.~\cite{KP} for details.} which is valid both for 
Dirac and Majorana fermions, has to be cancelled 
by $\delta u^{L,R}_{ii}$. 
The earlier UV divergence of $\delta u$ is 
consistent with the running of the mass matrix of $f$ 
in the gauge eigenbasis \cite{Urun1,Urun2,KP}. The renormalized mixing 
matrix $U$ is then given by specifying the finite part of $\delta u$. 

The modified minimal subtraction (\ms) scheme is simplest 
and proven to give gauge-independent renormalized 
parameters \cite{RGExi}. 
However, the decoupling of heavy particles is not manifest in this scheme. 
The cancellation of the dependence on the renormalization scale 
$Q$ between running parameters and different parts of the amplitude is 
often quite delicate and complicated. 
In addition, the ${\cal O}(1/(m_i^2-m_j^2))$ singularity 
for $m_i\simeq m_j$ remains in the amplitudes. 
These properties make the \ms scheme inconvenient in realistic studies. 
On the other hand, the renormalized mixing matrices may also be 
defined directly in terms of the physical observables. This method is 
manifestly independent of the gauge fixing and renormalization scale. 
However, the form of the counterterm strongly depends on the chosen 
observables and is often very complicated. 

It is therefore natural to investigate the method to define the 
renormalized mixing matrices which are independent of the 
renormalization scale and of the specific processes. 
In the study of the radiative correction to the CKM matrix, 
Denner and Sack \cite{DS} proposed to cancel the total anti-hermitian 
part of $\delta Z_{ij}$ by $\delta u$, choosing 
\beq
\delta u_{ij} = \frac{1}{4} (\delta Z_{ij} - \delta Z^*_{ji} ). \label{eq12}
\eeq
This is usually called the on-shell renormalization of 
the mixing matrix. Equation (\ref{eq5}) is then rewritten as 
\beq
U^{(0)*}_{j\alpha}\left( \delta_{ji} + \frac{1}{2}\delta Z_{ji}\right) 
= (U^{\rm OS})^*_{j\alpha}\left( 
\delta_{ji} + \frac{1}{4}(\delta Z_{ji} + \delta Z^*_{ij}) \right). 
\label{eq13}
\eeq
One important feature of Eq.~(\ref{eq13}) is that 
all ${\cal O}(1/(m_i^2-m_j^2))$ singularities in $\delta Z_{ij}$ are 
absorbed into the renormalized $U^{\rm OS}$. Also, $U^{\rm OS}$ is 
independent of the \ms renormalization scale. These properties are equally 
valid both for fermions and scalars. 

The mixing of quarks in different generations needs a special care 
for there is no unique ``gauge eigenbasis'' for them. 
Instead, one can discuss only the difference between the mixing of 
left-handed up-type quarks and that of 
down-type quarks, namely the CKM matrix $V_{ij}=(U^{u_L})_{i\alpha}
(U^{d_L})^*_{j\alpha}$. The counterterm for the on-shell CKM matrix 
is then given by \cite{DS,gaugedep} 
\beq
\delta V_{ij}=\delta u^{u_L}_{ik}V_{kj}+\delta u^{d_L*}_{jk}V_{ik}, 
\label{eq14}
\eeq
where $\delta u^{q_L}$ are given by Eq.~(\ref{eq12}). 

\section{Gauge dependence of wave function corrections 
and on-shell mixing matrices}

Since the proposal in Ref.~\cite{DS}, however, the dependence of 
the on-shell mixing matrix on the gauge fixing parameters $\xi$ 
has not been examined for a long time. 
Recent studies \cite{gaugedep,KMS,barroso} showed that the on-shell 
renormalization of the CKM matrix introduces gauge dependence into 
one-loop amplitudes for the $W^+\rightarrow u_i\bar{d}_j$ decays through 
the counterterm $\delta V_{ij}$. 
They proposed alternative definitions for quark mixing matrix which are 
independent of the renormalization scale. References \cite{gaugedep,KMS} 
used a modified process-independent definition for the CKM matrix. 
As shown in this section, 
their definition strongly relies on the gauge representation of quarks. 
Reference \cite{barroso} fixed the renormalized CKM matrix by using the 
amplitudes of the decays $W^+\rightarrow u_i\bar{d}_j$ 
(or $t\rightarrow W^+d_j$). To keep the renormalized CKM matrix unitary, 
four processes have to be selected out of nine possible ones. 
As a result, the forms of the corrected amplitudes become 
very asymmetric with respect to generation indices $(i,j)$. 
Thus, both methods cannot be directly applied 
for the renormalization of other mixing matrices. 
In this section we show another way to avoid the problem of 
gauge dependence of the on-shell scheme of Ref.~\cite{DS}. 

We first investigate the gauge parameter dependences of the wave function 
correction $\delta Z$ and of the counterterm $\delta u$ for 
the on-shell mixing matrix in general cases. 
We use the fact that, in the $R_\xi$ gauge, 
the dependence of the one-particle irreducible Green functions 
on the gauge parameters $\xi$ is controlled by the 
Nielsen identities \cite{nielsen1,PS}, a kind of 
the Slavnov-Taylor identities which follow from the extended 
Becchi-Rouet-Stora (BRS) symmetry \cite{PS} of the theory. 
The identity for the 
gauge parameter dependence of the inverse propagator $\Gamma_{ij}(p)$ 
for the transition $f_j\rightarrow f_i$ takes 
the following form \cite{nielsenSM} 
\begin{equation}
\partial_{\xi}\Gamma_{ij}(p) \equiv \partial\Gamma_{ij}(p)/\partial\xi = 
-\Gamma_{\chi\bar{f}_i K_l}(p)\Gamma_{lj}(p)
-\Gamma_{il}(p)\Gamma_{\chi K_{\bar{l}} f_j}(p).  \label{eq15}
\end{equation}
Here $\Gamma_{\chi\bar{f}_i K_l}(p)$ is the vertex function with 
$\bar{f}_i$, $\chi$, the ``BRS variation'' of the gauge 
parameter $\xi$ \cite{PS,nielsenSM}, and 
$K_l$, the source associated with the BRS variation 
of $f_l$. $\Gamma_{\chi K_{\bar{l}} f_j}(p)$ is its conjugate. 
Since the identity (\ref{eq15}) is determined by the form of the 
gauge-fixing terms \cite{nielsenSM}, it holds for general gauge 
theories in the $R_\xi$ gauge fixing. 
In Eq.~(\ref{eq15}) $f$'s are assumed to be physical fields with 
gauge-independent masses, not the would-be Nambu-Goldstone (NG) bosons, 
Fadeev-Popov ghosts, or longitudinal modes of gauge bosons. 
Under this condition 
$\Gamma_{\chi\bar{f}_i K_l}(p)$ has no tree level contribution. 
It is also required that the renormalization does not 
introduce additional gauge dependence \cite{nielsenSM}. 
Especially, the shift of the vacuum expectation values (VEVs) of 
Higgs bosons by tadpole graphs should be cancelled 
in a gauge-independent way. 

The gauge dependence of the one-loop two-point 
functions $\Sigma_{ij}(p)$ of fermions is, in the tree-level mass basis, 
derived from general result (\ref{eq15}) as 
\begin{equation}
\partial_{\xi}\Sigma_{ij}(p) = 
\Lambda_{ij}(p) (\psla - m_j) 
+ (\psla - m_i) \overline{\Lambda}_{ij}(p),  \label{eq16}
\end{equation}
where $\Lambda(p)$ and $\bar{\Lambda}(p)$ are some one-loop 
Dirac spinors. After the decomposition 
\beq
\Lambda_{ij}(p)=
\Lambda_{Lij}(p^2)\psla P_L + \Lambda_{Rij}(p^2)\psla P_R +
\Lambda_{DLij}(p^2) P_L + \Lambda_{DRij}(p^2) P_R,  \label{eq17}
\eeq
and similar one for $\overline{\Lambda}$, the $\xi$ dependence of the 
components of $\Sigma$ in Eq.~(\ref{eq7}) is \cite{nielsenSM} 
\bea
\partial_{\xi}\Sigma_{Lij} &=& 
-m_j\Lambda_{Lij}-m_i\overline{\Lambda}_{Lij}+\Lambda_{DRij}
+\overline{\Lambda}_{DLij}, \nonumber\\
\partial_{\xi}\Sigma_{Rij} &=& 
-m_j\Lambda_{Rij}-m_i\overline{\Lambda}_{Rij}+\Lambda_{DLij}
+\overline{\Lambda}_{DRij}, \nonumber\\
\partial_{\xi}\Sigma_{DLij} &=& 
p^2\Lambda_{Rij}+p^2\overline{\Lambda}_{Lij}-m_j\Lambda_{DLij}
-m_i\overline{\Lambda}_{DLij}, \nonumber\\
\partial_{\xi}\Sigma_{DRij} &=& 
p^2\Lambda_{Lij}+p^2\overline{\Lambda}_{Rij}-m_j\Lambda_{DRij}
-m_i\overline{\Lambda}_{DRij}. \label{eq18}
\eea
The relations 
\beq
\overline{\Lambda}_{Lij}=\Lambda^*_{Lji},\;\;
\overline{\Lambda}_{Rij}=\Lambda^*_{Rji},\;\;
\overline{\Lambda}_{DLij}=\Lambda^*_{DRji},\;\;
\overline{\Lambda}_{DRij}=\Lambda^*_{DLji},\;\; \label{eq19}
\eeq
follow from the hermiticity of the effective action. 

By substituting them into Eq.~(\ref{eq9}), 
we obtain \cite{gaugedep} for $i\neq j$ 
\beq
\frac{1}{2}\partial_{\xi}(\delta Z^L_{ij})=
-m_j\overline{\Lambda}_{Rij}(m_j^2)-\overline{\Lambda}_{DLij}(m_j^2), 
\label{eq20}
\eeq
and similar result for $\delta Z^R_{ij}$. 
As a result, the original definition of the 
on-shell renormalized fermion mixing 
matrices in Eq.~(\ref{eq12}) has gauge parameter dependence. 
Explicit calculation 
shows that the gauge dependence of the counterterm $\delta u^L_{ij}$ 
is equal to 
\beq
\frac{1}{4}\partial_{\xi}(\delta Z^L_{ij}-\delta Z^{L*}_{ji})= 
\frac{1}{2}\left[
-m_j\overline{\Lambda}_{Rij}(m_j^2)-\overline{\Lambda}_{DLij}(m_j^2)
+m_i\overline{\Lambda}^*_{Rji}(m_i^2)+\overline{\Lambda}^*_{DLji}(m_i^2)
\right]\, , \label{eq21}
\eeq
which does not vanish in general. This is also the case for 
$\delta u^R_{ij}$ and $\delta u_{ii}$. 

A remarkable fact in Eq.~(\ref{eq20}) is that the factor $1/(m_i^2-m_j^2)$, 
which characterizes the off-diagonal $\delta Z_{ij}$, is cancelled 
for the gauge dependence. 
This is expected from the gauge independence 
of the total amplitudes \cite{BRS} with gauge-independent renormalization 
of the couplings. Since the gauge dependence of Eq.~(\ref{eq20}) has to 
be cancelled by that from other parts of the amplitudes 
which do not have the factor $1/(m_i^2-m_j^2)$, the factor cannot 
remain in Eq.~(\ref{eq20}). 
Similar cancellation occurs in the gauge dependence 
of the diagonal part $\delta u^L_{ii}=-\delta u^R_{ii}$, which is equal to 
\beq
\frac{i}{2}\partial_{\xi}({\rm Im}\,\delta Z^L_{ii})= 
\frac{i}{2}{\rm Im}\left[
-m_i\overline{\Lambda}_{Rii}(m_i^2)+m_i\overline{\Lambda}_{Lii}(m_i^2)
+\overline{\Lambda}_{DRii}(m_i^2)-\overline{\Lambda}_{DLii}(m_i^2)
\right]\, . \label{eq21d}
\eeq
The factor $1/m_i$ in Eq.~(\ref{eq9a}), which characterizes 
${\rm Im}(\delta Z_{ii})$, is cancelled in Eq.~(\ref{eq21d}). 
Another important point is that Eqs.~(\ref{eq21}, \ref{eq21d}) 
are UV finite. 

The mixing matrices of the scalars can be analyzed in the similar way. 
The one-loop two-point function $\Pi_{ij}(p^2)$ for scalars in 
the tree-level mass basis obeys the relation \cite{nielsenSM} 
\begin{equation}
\partial_{\xi} \Pi_{ij}(p^2) = 
\Lambda_{ij}(p^2) (p^2 - m_j^2) 
+ (p^2 - m_i^2) \Lambda^*_{ji}(p^2), \label{eq22}
\end{equation}
from the Nielsen identity. We assume that there are no mixings with 
unphysical modes. By substitution we obtain for $i\neq j$ 
\beq
\frac{1}{2}\partial_{\xi}(\delta Z_{ij})= -\Lambda^*_{ji}(m_j^2). 
\label{eq23}
\eeq
The gauge dependence of the counterterm (\ref{eq12}) for the 
on-shell mixing matrix for scalars is therefore 
\beq
\partial_{\xi}(\delta u_{ij})= 
-\frac{1}{2}\left[ \Lambda^*_{ji}(m_j^2) - \Lambda_{ij}(m_i^2) \right]\, ,
\label{eq24}
\eeq
which is UV finite but does not cancel in general. 
However, the factor $1/(m_i^2-m_j^2)$ is again cancelled 
in Eq.~(\ref{eq24}). 

According to the earlier observation, we can define the gauge-independent 
one-loop on-shell mixing matrices for fermions and scalars as follows. 
First, we split gauge-dependent parts without the factor $1/(m_i^2-m_j^2)$ 
from $\delta Z_{ij}$ and regard them as parts of 
the corrections to the attached vertex. They are eventually 
cancelled by the gauge dependence of the vertex and other corrections. 
Second, we give the counterterms for mixing matrices in terms of the 
remaining, gauge-independent part of $\delta Z_{ij}$. This procedure 
gives the one-loop corrected amplitudes which are expressed in terms of 
the on-shell mixing matrices and manifestly gauge independent. 
Of course, the choice of the gauge-independent parts of 
$\delta Z_{ij}$ has arbitrariness. For example, we can regard 
the results in the $R_\xi$ gauge with a given $\xi$ 
as the gauge-independent parts. 

Here we propose a method to specify the gauge-invariant parts of 
$\delta Z_{ij}$, inspired by the 
pinch technique \cite{pinchold,pinch,pinchfermi} 
to define gauge-independent form factors for gauge bosons. 
We consider a general process with the external on-shell particle $f_j$ 
which is either a fermion or a scalar, with incoming momentum $p$. 
One source of the gauge dependence of $\delta Z_{ij}$ is 
the graph of Fig.~1(a). 
As pointed out in Refs.~\cite{pinchold,pinch}, the longitudinal 
part of the propagator of the (massive) gauge boson $A$ triggers 
the Ward identity at the vertex $\mu$ as 
\begin{equation}
\ksla = - (\psla-m_i) + (\ksla+\psla-M)  + (M-m_i), \label{eqWf}
\end{equation}
for fermions, or 
\begin{equation}
k^{\mu}(k+2p)_{\mu}= - (p^2-m_i^2) + ((k+p)^2-M^2) + (M^2-m_i^2), \label{eqWs}
\end{equation}
for scalars, respectively. The first two terms of 
Eqs.~(\ref{eqWf}, \ref{eqWs}) cancel the propagators of $f_i$ and 
of the intermediate particle $F$ with a mass $M$, respectively, 
and yield the contributions 
of Figs.~1(b, c) (pinching). The last terms of 
Eqs.~(\ref{eqWf}, \ref{eqWs}) are the effect of the spontaneous 
breaking of the gauge symmetry for $A$ and are proportional to the 
couplings to the associated NG boson. The part of Fig.~1(a) where 
the last terms are picked up at the vertex $\mu$ is further decomposed 
into three parts by the Ward identity (\ref{eqWf}, \ref{eqWs}) at $\nu$. 
The part which cancels $f_j$ propagator vanishes in on-shell amplitudes, 
while that which cancels $F$ propagator is included in the 
type of Fig.~1(c). The remaining part where 
the last terms of the Ward identity are picked up at both vertices 
does not fit into Figs.~1(b, c). 
To satisfy the Nielsen identities (\ref{eq16}, \ref{eq22}), 
this part has to be combined with the contribution from Fig.~1(d) 
by the NG boson $\phi_A$ to yield a gauge-independent sum. 
This result should be thus equal to the contribution of Fig.~1(d) 
in the $\xi=1$ gauge. 

The contribution of Fig.~1(b) is manifestly consistent with 
the Nielsen identity. In contrast, the remaining gauge-dependent part, 
Fig.~1(c), cannot satisfy the identity by itself because of its 
$p$ independence. This part has to be 
cancelled by the contributions from the Higgs VEV shift [Fig.~1(e)] 
by the loops of unphysical modes for $A$ and, in the case of scalars, 
by the ``seagull'' contributions with four-point 
couplings $f_i^*f_jA^{\mu}A_{\mu}$ [the same topology as Fig.~1(c)] 
and $f_i^*f_j\phi_A\phi_A$. 
Again, the result should be gauge-independent and therefore 
equal to the one in the $\xi=1$ gauge. 
We have verified that, for the cases discussed in Sections 4 and 5, 
the earlier cancellation of the gauge dependence really occurs and 
that the contribution of Fig.~1(b) is equal to the difference from the 
result in the $\xi=1$ gauge. 

It is then natural to identify the contribution of Fig.~1(b) to 
$\delta Z_{ij}$ as the gauge-dependent pinch term, in analogy to 
Ref.~\cite{pinch}, and to regard this as a part of the vertex 
corrections. 
Then, in this manner, we may regard the on-shell mixing 
matrices in the $\xi=1$ gauge as the gauge-independent ones. 
The cancellation of the UV divergence, renormalization scale dependence, 
and the ${\cal O}(1/(m_i^2-m_j^2))$ singularity is not 
affected by this modification of the original definition of 
the on-shell mixing matrices. 
Note that the agreement of the $\xi=1$ and the 
pinch technique results has been observed for 
the QCD correction to the off-shell quark propagator \cite{pinchfermi}. 
Note also that we have not considered, for scalars, the 
possible trigger of the Ward identity (\ref{eqWs}) at the vertex $\mu$ 
by the momentum $(k+2p)_{\nu}$ at the vertex $\nu$ in Fig.~1(a), 
which was done for the couplings of the gauge and NG bosons \cite{pinch} 
to satisfy the Ward identities among corrected vertices. 

We finally comment on other definitions for the UV finite 
and process-independent mixing matrices for fermions. 
As the first example, Ref.~\cite{nojiri} proposed 
a definition of the UV finite and 
momentum-dependent effective mixing matrices 
[$\overline{U}^L(p^2)$, $\overline{U}^R(p^2)$] for fermions. 
The counterterms for $\overline{U}$ are given by, 
instead of Eq.~(\ref{eq9}), 
\bea
\delta\bar{u}^L_{ij}(p^2)&=&\frac{1}{m_i^2-m_j^2}\left[
\frac{1}{2}(m_i^2+m_j^2)\Sigma_{Lij}(p^2) \right. \nonumber\\
&& \left. + m_im_j\Sigma_{Rij}(p^2) 
+m_i\Sigma_{DLij}(p^2)+m_j\Sigma_{DRij}(p^2) \right]~, \label{eq25}
\eea
and similar form for $\delta\bar{u}^R_{ij}(p^2)$. 
Similar to the on-shell $U$ by Ref.~\cite{DS}, $\overline{U}(p^2)$ absorb 
the ${\cal O}(1/(m_i^2-m_j^2))$ singularity when the couplings of $f_i$ 
are expressed in terms of $\overline{U}_{i\alpha}(p^2=m_i^2)$. 
Unfortunately, this definition also shows gauge 
dependence. From Eq.~(\ref{eq18}) we obtain 
\beq
\partial_{\xi}[\delta\bar{u}^L_{ij}(m_i^2)]=
\frac{1}{2}\left( m_j\Lambda_{Lij}+m_i\overline{\Lambda}_{Lij}
+2m_i\Lambda_{Rij}+\Lambda_{DRij}-\overline{\Lambda}_{DLij}
\right)(p^2=m_i^2). \label{eq26}
\eeq
In Eq.~(\ref{eq26}) the factor $1/(m_i^2-m_j^2)$ is again 
cancelled. To avoid this gauge dependence, the gauge-dependent term 
of the self energy (\ref{eq7}) has to be rearranged such that 
it vanishes in the counterterm (\ref{eq25}). 
As the second example, Ref.~\cite{gaugedep} proposed to 
renormalize the CKM matrix in terms of the zero-momentum self 
energies for quarks. The counterterm is then 
\beq
\delta u^L_{ij}({\rm [7]})=\frac{1}{m_i^2-m_j^2}\left[
\frac{1}{2}(m_i^2+m_j^2)\Sigma_{Lij}(0) 
+m_i\Sigma_{DLij}(0)+m_j\Sigma_{DRij}(0) \right]~.  \label{eq27}
\eeq
This definition gives the renormalized CKM matrix which is 
gauge-independent and UV finite. 
However, its validity relies on the fact that quark 
couplings to $W^{\pm}$ are purely left-handed. Moreover, Eq.~(\ref{eq27}) 
does not absorb the ${\cal O}(1/(m_i^2-m_j^2))$ singularity. 
Thus, this definition has to be greatly modified for the renormalization 
of other mixing matrices. 

\section{CKM matrix: example for fermion mixing}

In this and the next sections we show the explicit form of the gauge 
dependence of the on-shell mixing matrices, both for fermions and 
for scalars. In this section we discuss the on-shell CKM matrix, 
following previous studies \cite{gaugedep,KMS}. 

The off-diagonal parts of the one-loop self energies $\Sigma^q_{ij}(p)$ 
of the quarks receive gauge-dependent contribution from the 
$W^{\pm}$ loops \cite{sirlin,DS}. 
The $\xi_W$ dependent part of $\Sigma^u_{ij}(p)$ for 
up-type quarks $u_i=(u, c, t)$, 
namely the difference from the result in the $\xi_W=1$ gauge, 
takes the following form; 
\bea
\Sigma^u_{ij}(p)|_{\xi_W} &=& (1-\xi_W)\frac{g_2^2}{32\pi^2}\sum_k 
V_{ik}V^*_{jk}
\left\{ (\psla-m_{u_i}) \beta^{(1)}_{Wd_k}(p^2)\psla P_R (\psla-m_{u_j}) 
\right.\nonumber\\
&& -(\psla-m_{u_i}) P_L 
\left[m_{d_k}^2\beta^{(0)}_{Wd_k}(p^2)
-m_{u_j}\beta^{(1)}_{Wd_k}(p^2) \psla +\frac{1}{2}\alpha_W 
\right] \nonumber \\
&& \left. -\left[ 
m_{d_k}^2\beta^{(0)}_{Wd_k}(p^2)-m_{u_i}\beta^{(1)}_{Wd_k}(p^2)\psla 
+\frac{1}{2}\alpha_W \right] P_R (\psla-m_{u_j}) \right\}\;. \label{eq28}
\eea
Here we define 
\bea
\frac{i}{16\pi^2}\alpha_i &=& 
\int \frac{d^nq}{(2\pi)^n}\frac{1}{(q^2-m_i^2)(q^2-\xi_i m_i^2)}, \\
\frac{i}{16\pi^2}\beta^{(0)}_{ij}(p^2) &=& 
\int \frac{d^4q}{(2\pi)^4}
\frac{1}{(q^2-m_i^2)(q^2-\xi_i m_i^2)((q+p)^2-m_j^2)}, \\
\frac{i}{16\pi^2}\beta^{(1)}_{ij}(p^2)p_{\mu} &=& 
\int \frac{d^4q}{(2\pi)^4}
\frac{(q+p)_{\mu}}{(q^2-m_i^2)(q^2-\xi_i m_i^2)((q+p)^2-m_j^2)}, 
\eea
where $n=4-2\epsilon$. 
$\Sigma^d_{ij}(p)|_{\xi_W}$ for down-type quarks $d_i=(d, s, b)$ is obtained 
by replacing ($u_i$, $u_j$, $d_k$, $V_{ik}V^*_{jk}$) in Eq.~(\ref{eq28}) by 
($d_i$, $d_j$, $u_k$, $V^*_{ki}V_{kj}$). 
Equation (\ref{eq28}) is equivalent to the results in 
Refs.~\cite{KMS,nielsenSM}, 
except that Eq.~(\ref{eq28}) includes the gauge-dependent part of 
the Higgs VEV shift in $\Sigma^u_{ii}$, 
by tadpoles with $W^{\pm}$ and associated unphysical modes. 
This corresponds to defining renormalized Higgs VEV as the 
minimum of the tree-level potential \cite{degrassi,mfermion,nielsenSM}, 
which is gauge-independent in the \ms scheme. 
By the addition of the Higgs VEV shift, 
Eq.~(\ref{eq28}) manifestly satisfies the Nielsen identity (\ref{eq16}). 
Instead, one may also add the counterterms for pole masses of 
quarks to the diagonal elements to satisfy Eq.~(\ref{eq16}). 
This difference does not affect the present discussion. 

The counterterm for the on-shell CKM matrix in the original 
definition \cite{DS}, without separating Eq.~(\ref{eq28}), 
has gauge dependence as 
\beq
(\delta V_{ij})_{\xi}=X^u_{ik}V_{kj}+V_{il}X^d_{jl}. \label{eq32}
\eeq
$X^u_{ik}$ ($i\neq k$) is obtained from Eq.~(\ref{eq28}) as 
\bea
X^u_{ik}&=&(1-\xi_W)\frac{g_2^2}{64\pi^2}V_{il}V_{kl}^*
\left[ -m_{u_k}^2\beta^{(1)}_{Wd_l}(m_{u_k}^2) 
+ m_{u_i}^2\beta^{(1)}_{Wd_l}(m_{u_i}^2) \right. \nonumber \\
&& \left. + m_{d_l}^2\beta^{(0)}_{Wd_l}(m_{u_k}^2) 
- m_{d_l}^2\beta^{(0)}_{Wd_l}(m_{u_i}^2)
\right]. \label{eq33}
\eea
$X^d_{jl}$ has a similar form. 
Equation (\ref{eq32}) causes gauge-dependent amplitudes 
for the $W\bar{u}_id_j$ interactions \cite{gaugedep,KMS,barroso}. 
Numerically, Eq.~(\ref{eq32}) is greatly suppressed, partly by the 
Glashow-Iliopoulos-Maiani (GIM) mechanism \cite{GIM}, and 
completely negligible in practice \cite{DS}. 
The relative corrections are largest to ($V_{cb}$, $V_{ub}$, $V_{td}$, 
$V_{ts}$), but are at most ${\cal O}(10^{-6})$. 
Nevertheless, this is not satisfactory for theoretical point of view. 
The study in previous section 
shows, however, that one can give the counterterm $\delta V$ in terms of 
$\xi=1$ parts of $\Sigma_{ij}^{u}$ and $\Sigma_{ij}^{d}$. The original 
calculation in Ref.~\cite{DS} is thus interpreted as 
a gauge-independent one after the rearrangement. 

\section{Left-right mixing of squarks: example for scalar mixing} 

We next consider the renormalization of the left-right mixing of squarks 
in the MSSM, for an example for the mixing of scalar particles. 
For simplicity, we treat the mixing of two eigenstates of 
the top squarks, ignoring CP violation and mixing with different 
generations. 

The gauge eigenstates ($\sq_L$, $\sq_R$) of squarks, which are 
the superpartners of a quark $q$, mix with each other by 
spontaneous breaking of SU(2)$\times$U(1) gauge symmetry \cite{mssm,gh}. 
Their mass eigenstates $\sq_i(i=1,2)$ are related to the 
gauge eigenstates $\sq_{\alpha}(\alpha=L,R)$ by 
$\sq_i=R^{\sq}_{i\alpha}\sq_{\alpha}$ with the left-right mixing matrix 
\beq 
R^{\sq}_{i\alpha}=\left(
\begin{array}{cc}\cos\theta_\sq & \sin\theta_\sq \\
-\sin\theta_\sq & \cos\theta_\sq \end{array}\right)~. \label{eq34}
\eeq

The renormalization of the squark sector is often performed by 
specifying the poles masses of ($\sq_1$, $\sq_2$) and the mixing angle 
$\theta_{\sq}$, as in Refs.~\cite{ACD,eesqsq,tstnt,sqchnt,beenakker3,
hsq,hsq2,sqyukawa,sqhx,improved}. 
Following the result in section 2, the counterterm 
$\delta\theta_{\sq}$ is given by \cite{sqyukawa,improved} 
\beq
\delta\theta_{\sq} = \delta r_{12} =
\frac{1}{2(m_{\sq_1}^2 - m_{\sq_2}^2)} 
\left[ \Pi^\sq_{12}(m_{\sq_1}^2)+\Pi^\sq_{12}(m_{\sq_2}^2) \right]\, ,
\label{eq35}
\eeq
with $\Pi^\sq_{12}(p^2)$ the off-diagonal self energy of squarks 
in the tree-level mass basis. Although many other on-shell 
definitions \cite{ACD,eesqsq,tstnt,sqchnt,beenakker3,hsq,hsq2} have 
been used in the studies of the SUSY QCD corrections, 
they are either unable to be applied for other loop corrections, or 
too specific for the squark processes considered there. 

We consider the on-shell mixing matrix for top squarks $\st_i$. 
The gauge-dependent part of the unrenormalized two-point 
function $\Pi^{\st}_{ij}(q^2)$, namely the difference from the 
results in the $\xi_Z=\xi_W=1$ gauge \cite{pierce}, takes the following form: 
\bea
\Pi^{\st}_{ij}(p^2)|_{\xi}&=& 
\frac{g_Z^2}{16\pi^2}(1-\xi_Z)\sum_k
(R^{\st}_{i1}R^{\st}_{k1}T_{3t}-\delta_{ik}s_W^2Q_t)
(R^{\st}_{k1}R^{\st}_{j1}T_{3t}-\delta_{kj}s_W^2Q_t) \nonumber\\
&& \times\left[ 
-\frac{1}{2}(2p^2-m_{\st_i}^2-m_{\st_j}^2)\alpha_Z 
+\left\{ (p^2-m_{\st_i}^2)(p^2-m_{\st_j}^2) \right. \right. \nonumber\\
&& \left.\left. +(p^2-m_{\st_i}^2)(m_{\st_j}^2-m_{\st_k}^2) 
+(m_{\st_i}^2-m_{\st_k}^2)(p^2-m_{\st_j}^2) \right\}
\beta^{(0)}_{Z\st_k}(p^2) \right] \nonumber\\
&& +\frac{g_2^2}{32\pi^2}(1-\xi_W)R^{\st}_{i1}R^{\st}_{j1}\left[
-\frac{1}{2}(2p^2-m_{\st_i}^2-m_{\st_j}^2)\alpha_W \right.\nonumber\\
&& +\sum_k(R^{\sb}_{k1})^2
\left\{ (p^2-m_{\st_i}^2)(p^2-m_{\st_j}^2)
+(p^2-m_{\st_i}^2)(m_{\st_j}^2-m_{\sb_k}^2) 
\right.\nonumber\\
&& \left. \left. +(m_{\st_i}^2-m_{\sb_k}^2)(p^2-m_{\st_j}^2) \right\}
\beta^{(0)}_{W\sb_k}(p^2)  \right]\, . \label{eq36}
\eea
Here $T_{3t}=1/2$, $Q_t=2/3$, and $s_W^2=\sin^2\theta_W$. 
As before, Eq.~(\ref{eq36}) includes the gauge-dependent shifts of the 
two Higgs VEVs for gauge-independent renormalization of the VEVs. 
In contrast to the SM case, they also contribute to the $i\neq j$ parts. 
The result (\ref{eq36}) satisfies the Nielsen identity (\ref{eq22}). 

The magnitude of the gauge dependence of the on-shell 
$\delta\theta_{\st}$ is very sensitive to squark parameters. 
For a parameter choice ($M_{\tilde{Q}}$, $M_{\tilde{U}}$, 
$M_{\tilde{D}}$)$=$(350, 300, 400)~GeV, $\tan\beta=4$, 
($\mu$, $A_t$, $A_b$)$=$($-400$, 300, 0)~GeV, and $0<\xi<10$, 
$\xi_W$ and $\xi_Z$ dependent parts of $\delta\theta_{\st}$ 
may be as large as 0.008 and 0.003, respectively. 
Although too small for realistic phenomenology, 
they are much larger than the $\xi_W$ dependence of the on-shell CKM matrix. 
This is partly due to the absence of the GIM cancellation, following from 
that $\st_L$ and $\st_R$ have different gauge representations. 
As is already shown, these gauge dependence of $\theta_{\st}$ can 
be avoided by removing the contribution of Eq.~(\ref{eq36}) from 
off-diagonal wave function corrections $\delta Z_{12}$ for top squarks, 
cancelling it by other gauge dependences of the amplitude, 
and then giving $\delta\theta_{\st}$ by the remaining part 
of $\delta Z_{12}$. 

\section{Conclusion}

In this paper, we investigated the gauge parameter dependence of 
the on-shell renormalized mixing matrices for scalars and fermions 
at the one-loop level. 
It has been shown recently that the on-shell renormalization 
of the CKM matrix in the definition by Ref.~\cite{DS} is gauge dependent. 
By using the Nielsen identities for self energies, 
we demonstrated that this gauge dependence exists for the 
on-shell mixing matrices in general cases. We also showed 
that this gauge dependence can be avoided by the 
following procedure; split the gauge-dependent parts from 
the off-diagonal wave function corrections in the manner similar to the 
pinch technique, and then giving the counterterm for the mixing matix 
in terms of the remaining, gauge-independent parts. 
The subtraction of the UV divergence and ${\cal O}(1/(m_i^2-m_j^2))$ 
singularity is not affected by this modification. 
The on-shell scheme in Ref.~\cite{DS} in the $\xi=1$ gauge can be then 
regarded as gauge-independent one. 
Finally, we presented explicit calculation of the 
gauge-dependence of the mixing matrices in two cases, 
CKM matrix and left-right mixing of squarks, and verified 
the result from the Nielsen identities. 

We did not treat the mixings of the gauge bosons and of the Higgs bosons. 
In principle, our method would also be applicable for these mixings. 
When applied for the mixing of the gauge bosons $\gamma$ and $Z$, 
the square of the renormalized mixing angle $\sin^2\theta_W({\rm OS})$ 
agrees with the effective angle $s_*^2(m_Z^2)$ defined 
in Ref.~\cite{LEPew}, at the one-loop level. But the inclusion 
of the absorptive part of the $Z$ boson propagator is necessary 
for realistic studies. 
The correction to the mixing of the MSSM Higgs bosons in diagrammatic 
calculation \cite{cpr,dabel,higgs2l} is a very interesting subject. 
However, due to the mixing of physical Higgs bosons with 
unphysical modes, a separate consideration is necessary. 
We expect to study the case of the MSSM Higgs bosons in future. 

\section*{Acknowledgements}

The author thanks Helmut Eberl and Wolfgang Hollik for useful discussions. 
This work was supported in part by the Grant-in-aid for 
Scientific Research Nos.~12740131 and 12047201. 


\newpage

\begin{figure}[htbp]
\begin{center}
\includegraphics[width=15cm]{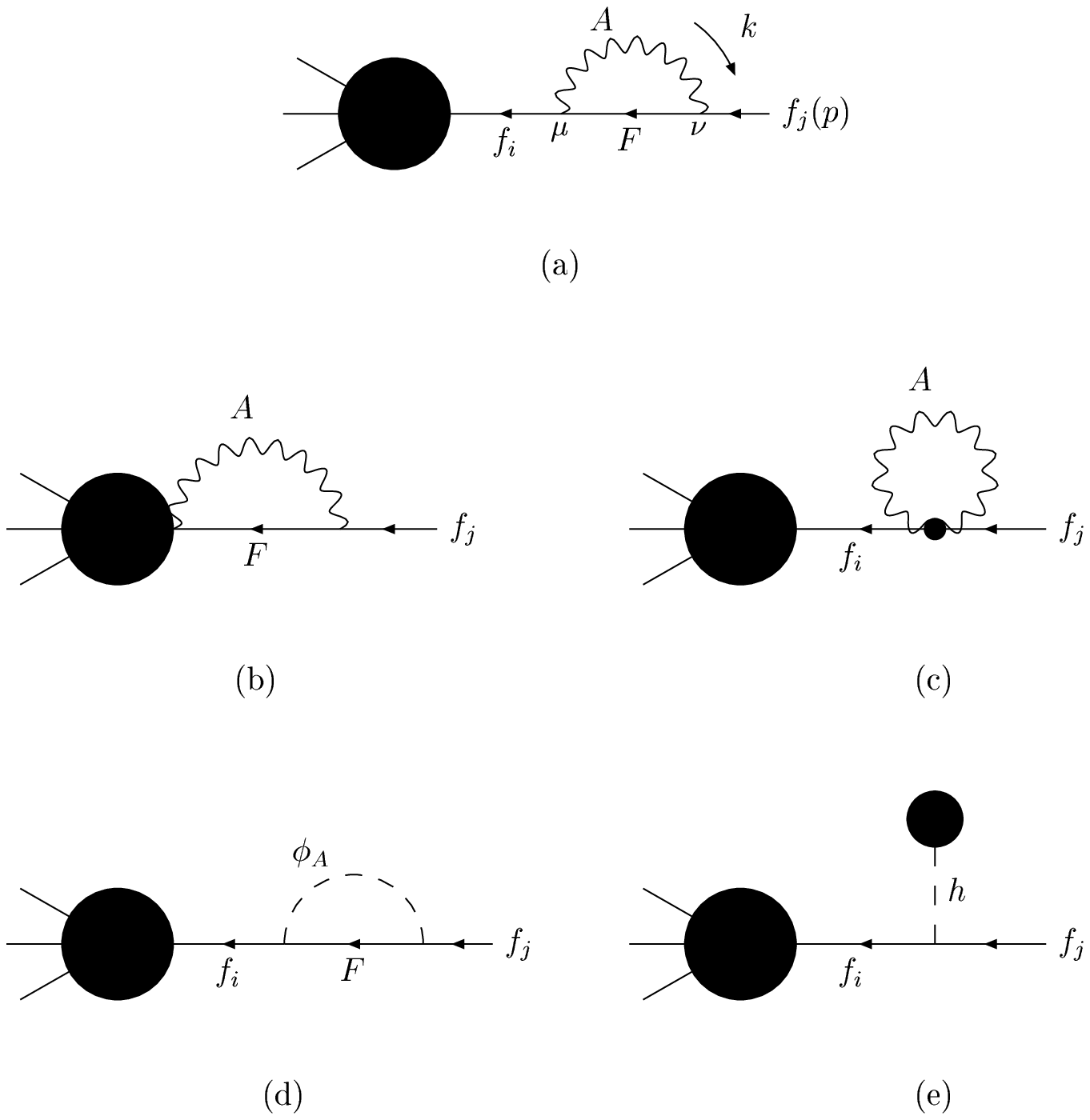}
\end{center}
\caption{ 
The gauge-dependent contributions to $\delta Z_{ij}$ for 
a general process with external on-shell $f_j$, 
which is either a fermion or a scalar, 
from the loops of massive gauge boson $A$ and intermediate particle $F$. 
Graphs (b, c) are the ``pinch terms'' stemming from (a). 
Graph (d) is a contribution of the NG boson $\phi_A$ associated with $A$. 
Graph (e) represents the shift of the VEV of Higgs bosons $h$ by the 
loops of $A$, $\phi_A$, and Fadeev-Popov ghosts. Inclusion of (e) is 
necessary for gauge-independent renormalization of the Higgs VEVs. 
}
\label{fig1}
\end{figure}

\end{document}